\documentclass[%prl,
aps,%preprint %tightenlines
 ,twocolumn
]{revtex4}
\usepackage{graphicx}

\begin{document}

\title{Photon Sorting, Efficient Bell Measurements and a Deterministic CZ Gate using a Passive Two-level Nonlinearity}

\author{T.C.Ralph$^1$, I.~S\"{o}llner$^2$, S.~Mahmoodian$^2$, A.G.White$^1$ and P.~Lodahl$^2$}\affiliation{$^1$Centre for Quantum Computation and Communication Technology, \\
School of Mathematics and Physics, University
of Queensland, Brisbane, Queensland 4072, Australia \\
$^2$ Niels Bohr Institute, University of Copenhagen, Blegdamsvej 17, DK-2100 Copenhagen, Denmark}

\date{\today}
\begin{abstract}
{Although the strengths of optical non-linearities available experimentally have been rapidly increasing in recent years, significant challenges remain to using such non-linearities to produce useful quantum devices such as efficient optical Bell state analysers or universal quantum optical gates. Here we describe a new  approach that avoids the current limitations by combining strong non-linearities with active Gaussian operations in efficient protocols for Bell state analysers and Controlled-Sign gates. }

\end{abstract}

%
%\pacs{03.67.Dd, 42.50.Dv, 89.70.+c}

\maketitle

\vspace{10 mm}
{\it Introduction}:
It has long been the dream of the quantum optics community to use non-linear optical interactions to produce deterministic quantum logic operations, such as a Controlled-Sign (CZ) gate, between individual photons \cite{MIL89}. In combination with easily implemented single qubit operations the CZ gate produces a universal set of quantum logic operations that would enable applications from quantum repeaters to full scale quantum computation with photons. Unfortunately not only is it very difficult to achieve the strengths of non-linearity required for such gates, but it has been predicted for several candidate systems that strong non-linearities inevitably add noise and/or distort the optical modes of the single photons sufficiently that successful operation, even under ideal conditions, is impossible \cite{SHA06, SHE07, XU13}. 

One such candidate system with various possible physical implementations is a single two-level emitter deterministically coupled to a one-dimensional photonic waveguide. We will refer to such a system here as a Two-Level Scatterer (TLS). One interesting capability of such systems is to separate the single and two photon components of an optical mode into two separate modes. This has been referred to as photon sorting \cite{WIT12}. In principle photon sorting, if efficient and mode preserving, could be used to perform full Bell measurements, and to implement deterministic quantum logic gates between, photonic qubits. However, it has been shown that the TLS introduces mode distortion between the single and two photon components in the form of spectral entanglement of the two photon component \cite{SHE07} which has been argued to be unavoidable \cite{XU13}. As a result photon sorting is inefficient. Whilst it has been shown that by combining multiple interactions with a TLS with linear optics it is possible to perform near deterministic Bell measurements the proposed scheme requires $80$ separate scattering events to obtain a probability of success of about 95\% \cite{WIT12}.

Here we show that by adding active Gaussian optics to our tool-box of scatterer plus passive linear optics we are able to perform a deterministic Bell measurement using only 4 interactions with a TLS. Similarly, in principle it becomes possible to implement deterministic quantum logic gates in this way. Ironically, it is by exploiting the inherent mode distortion of the scatterer that these operations become possible.

{\it Action of the Scatterer}:
We consider a TLS formed by placing a two-level emitter in a nanophotonic cavity or waveguide that is designed for unidirectional interaction \cite{LUK14, RAUSCH14, SOL14}, cf. Fig.\ref{e1}(a). The monochromatic creation operator for the input mode, with wave number $k$, is scattered such that the corresponding creation operator for the output mode is given by \cite{SHE05}
\begin{eqnarray}
\hat a_{k, in}^{\dagger} \to t_k \hat a_{k, out}^{\dagger}
\label{e1}
\end{eqnarray}
where
\begin{eqnarray}
t_k = {{c k - \omega_0+i(\gamma-\Gamma)/2}\over{c k - \omega_0+i(\gamma+\Gamma)/2}}
\label{e2}
\end{eqnarray}
with $c$ the speed of light, $\omega_0$ the resonant frequency of the two-level emitter, $\Gamma$ the coupling strength between the emitter and a uni-directional waveguide mode and $\gamma$, the coupling strength of the emitter to modes other than the directional waveguide mode of interest. Eq.\ref{e1} describes a linear transformation similar to that produced by reflection from a single ended optical cavity. In general the output state will be mixed due to losses into other modes and the relevant figure of merit is the directional $\beta_{dir}$-factor defined as $\beta_{dir} = \Gamma/(\gamma + \Gamma)$ \cite{SOL14}. In the ideal case for which losses are negligible, i.e. $\gamma = 0$, the scattering is unitary and we can write the input-output relation for a single photon state with an arbitrary pulse shape, $f(k)$, as
\begin{eqnarray}
|1_f \rangle &=& \int dk f(k) \hat a_{k, in}^{\dagger} |0 \rangle  \nonumber \\
&\to& |1_{f'} \rangle = \int dk  f(k) t_k \hat a_{k, out}^{\dagger} |0 \rangle
\label{e3}
\end{eqnarray}

\begin{figure}[htb]
\begin{center}
\includegraphics*[width=8cm]{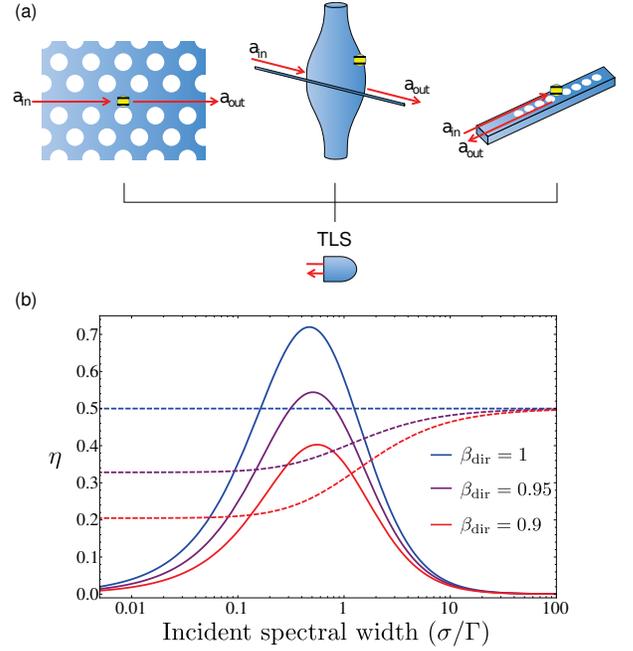}
\caption{(a) Possible physical implementations of an efficient TLS exploiting unidirectional coupling obtained by implementing a two-level emitter in a chiral photonic-crystal waveguide (left) \cite{SOL14}, a whispering-gallery resonator (middle) \cite{RAUSCH14}, or a photonic-crystal cavity (right) \cite{LUK14}. (b) $\eta$ as a function of spectral width of the incoming pulse in the case of a Lorentzian spectral response and for three different values of the $\beta_{dir}$-factor ($\beta_{dir} =1$ is the top solid line). The dashed lines show ${\epsilon_1^2 \over 2}$ for the different values of $\beta_{dir}$. Where the solid and corresponding dashed lines meet are the optimal operation points. In order to minimize loss, the crossing points at higher spectral widths should be chosen as the operation points.}
\label{figcalcs}
\end{center}
\end{figure}

We now consider two photon inputs. The equivalent of Eq.(\ref{e1}) for a pair of monochromatic creation operator with wave numbers $k_1$ and $k_2$ is \cite{SHE07, SHE07a}
\begin{eqnarray}
\label{e4}
\hat a_{k_1, in}^{\dagger} \hat a_{k_2, in}^{\dagger} &\to &
t_{k_1} \; \hat a_{k_1, out}^{\dagger} \; t_{k_2}  \; \hat a_{k_2, out}^{\dagger} \nonumber \\
&+& T_{k_1,k_2,p_1,p_2} \; \hat a_{p_1, out}^{\dagger} \; \hat a_{p_2, out}^{\dagger}
\label{e4}
\end{eqnarray}
where
\begin{eqnarray}
T_{k_1,k_2,p_1,p_2}  = {{i \sqrt{\Gamma}}\over{2 \pi}} \delta(k_1 + k_2 - p_1 - p_2) s_{p_1} s_{p_2} (s_{k_1} + s_{k_2}) \nonumber \\
\label{e5}
\end{eqnarray}
with
%
%\begin{eqnarray}
$s_{k} = {{1}\over{i \sqrt{\Gamma}}}(1-t_k)$.
%\label{e6}
%\end{eqnarray}
%
Eq.\ref{e4} describes a highly non-linear interaction which produces entanglement between the spectral components of the two input photons. Again considering the ideal case for which $\gamma = 0$ we can write
\begin{eqnarray}
|2_f \rangle  \to |2_{f'} \rangle + |2_{fb} \rangle
\label{e7}
\end{eqnarray}
where
\begin{eqnarray}
|2_{f'} \rangle = \int dk_1 \; dk_2 f(k_1) t_{k_1} \hat a_{k_1, out}^{\dagger} f(k_2) t_{k_2} \hat a_{k_2, out}^{\dagger} |0 \rangle \nonumber \\
\label{e8}
\end{eqnarray}
and
\begin{eqnarray}
|2_{fb} \rangle &=& \int dk_1 \; dk_2 \; dp_1 \; dp_2 \; T_{k_1,k_2,p_1,p_2} \nonumber \\
&\times& f(k_1) \hat a_{p_1, out}^{\dagger} f(k_2) \hat a_{p_2, out}^{\dagger} |0 \rangle
\label{e9}
\end{eqnarray}
The solution in the form of Eq.\ref{e7} was presented in \cite{WIT12}. However, it is clear from the normalization of Eq.\ref{e7} that the states of Eq.\ref{e8} and (Eq.\ref{e9}) are not orthogonal. Improved physical insight into the process can be obtained by rewriting Eq.\ref{e7} in terms of orthogonal states. We obtain
\begin{eqnarray}
|2_f \rangle  \to (1-2 \eta) |2_{f'} \rangle + 2\sqrt{\eta(1-\eta)} |\bar 2_{f'} \rangle
\label{e10}
\end{eqnarray}
where
\begin{eqnarray}
\eta = {{1}\over{2}} |\langle 2_{f'} |2_{fb} \rangle|
\label{e11}
\end{eqnarray}
and $|\bar 2_{f'} \rangle$ is a normalized state satisfying $\langle 2_{f'}|\bar 2_{f'} \rangle = 0$. The value of $\eta$ depends on the specific pulse shape chosen for the input state. It can be calculated analytically for pulses with a Lorentzian spectral shape and is found to be
\begin{eqnarray}
\eta =\frac{4 \Gamma^2 \sigma  \left(3 \Gamma^2+38 \Gamma  \sigma +96 \sigma^2\right)}{(\Gamma +2\sigma )^3 (3 \Gamma +2\sigma) (\Gamma +6\sigma)}
\label{e12}
\end{eqnarray}
%
%\begin{figure}[htb]
%\begin{center}
%\includegraphics*[width=8cm]{figcalcs.pdf}
%\caption{(a) Possible physical implementations of an efficient TLS exploiting unidirectional coupling obtained by coupling a two-level emitter to a chiral photonic-crystal waveguide (left) \cite{SOL14}, a whispering-gallery resonator (middle) \cite{RAUSCH14}, or a photonic-crystal cavity (right) \cite{LUK14}. (b) $\eta$ as a function of spectral width of the incoming pulse in the case of a Lorentzian spectral response and for three different values of the $\beta_{dir}$-factor. \is{The optimal operation point corresponds to $\eta = 0.5.$}}
%\label{figcalcs}
%\end{center}
%\end{figure}
%
where $\sigma$ is the width of the Lorentzian. A plot of the behaviour of Eq.\ref{e12} as a function of $\sigma$ is shown in Fig.\ref{figcalcs}(b). If it was possible to achieve $\eta = 1$ then one could directly use two TLSs to build a deterministic CZ gate as the transformation would essentially be a Non-linear Sign shift (NS) gate - imposing a phase flip on the two photon component but not the single photon component of the state \cite{KNI01}. It would also be possible to achieve deterministic photon sorting via the scheme in Ref. \cite{WIT12}. Unfortunately, numerically it appears that $\eta$ is bounded by $0 \le \eta < 0.75$.
\begin{figure}[htb]
\begin{center}
\includegraphics*[width=8.5cm]{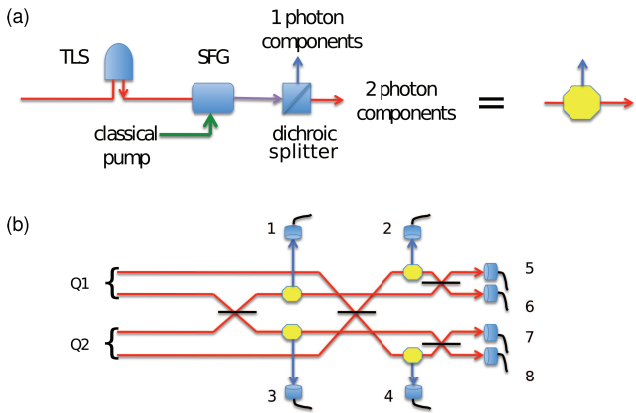}
\caption{Components of the photon sorter and Bell measurement device: (a) the photon sorter is constructed from a TLS followed by Sum Frequency Generation (SFG), where the classical pump is in the mode $f'$, followed by a dichroic beamsplitter which separates the single photon component at the sum frequency from the two photon component at the original frequency; (b) a Bell measurement can be implemented with linear optics and 4 photon sorters as shown. The four Bell states are unambiguously determined by the measurement of photons at particular detector combinations. In particular, $| \psi^{+} \rangle$: 1, 4 or 3, 2; $ | \psi^{-} \rangle$: 1, 2 or 3, 4; $| \phi^{+} \rangle$: 5, 8 or 6, 7; $ | \phi^{-} \rangle$: 5, 7 or 6, 8.}
\label{fig1}
\end{center}
\end{figure}

{\it Efficient Photon Sorting}:
One solution to this problem is to operate instead with $\eta = 0.5$, which is easily achieved with either a Lorentzian (see Fig.\ref{figcalcs}(b)) or Gaussian \cite{WIT12} mode function . An arbitrary superposition of single and two photon components is then transformed by the TLS as
\begin{eqnarray}
\alpha |1_f \rangle + \xi |2_f \rangle  \to \alpha |1_{f'} \rangle + \xi |\bar 2_{f'} \rangle
\label{e13}
\end{eqnarray}
With this choice of parameters the one and two photon components are completely mapped into co-propagating, but orthogonal, spatio-temporal modes which can in principle be perfectly separated with Gaussian transformations. This conclusion continues to be true for $\beta_{dir} < 1$ as shown in Fig.~\ref{figcalcs}b. Now the matching condition is $\eta = {{\epsilon_1^2}\over{2}}$ where $\epsilon_1$ is the probability that a single photon is scattered into the output mode by the TLS (see Supplementary Material for details).

Because the modes have overlapping spectral and temporal domains passive filtering will not be sufficient to perfectly separate them - instead active filtering is required. In particular, consider sum frequency generation (SFG). It was shown in Ref. \cite{ECK11} that by using suitably engineered SFG a quantum pulse gate can be produced which can efficiently extract a particular spatio-temporal mode from a multi-mode field. This works by choosing the pump field to perfectly match the spatio-temporal mode to be extracted. After interaction with a $\chi^{(2)}$ non-linear crystal it is this, and only this mode that is converted to the sum frequency. A passive frequency filter will then suffice to split the field into separate beams. The procedure is shown diagrammatically in Fig.\ref{fig1}(a) and can be represented mathematically as
\begin{eqnarray}
(\alpha |1_f \rangle + \xi |2_f \rangle) |0 \rangle_a  &_{\to}^{TLS}& (\alpha |1_{f'} \rangle + \xi |\bar 2_{f'} \rangle) |0 \rangle_a \nonumber\\
&_{\to}^{SFG}& \alpha |0 \rangle |1_{f'} \rangle_a + \xi |\bar 2_{f'} \rangle |0 \rangle_a
\label{e14}
\end{eqnarray}
where the mode function of the classical pump beam for the SFG is $t_k f(k)$ and the ket (initially in the vacuum state) labelled with the subscript $a$ is an ancilla mode at the sum frequency. Hence, using a single TLS plus sum frequency generation and passive filtering it is possible to produce a deterministic photon sorter. One should compare this with Ref.\cite{WIT12} where (assisted by only linear optics) 10 TLSs (or perhaps 10 interactions with a single TLS) are required to achieve, in principle, 95\% separation of the one and two photon components.

{\it Bell Measurement}:
Equipped with a deterministic photon sorter it is straightforward to construct a circuit from passive linear optics that can implement deterministic Bell measurements on dual rail single photon qubits. A dual rail qubit is where the logical value of the qubit is determined by which of two orthogonal modes is occupied, i.e. $|0 \rangle_L = |1 \rangle_u |0 \rangle_l$ and $|1 \rangle_L = |0 \rangle_u |1 \rangle_l$, where number kets for the two modes are labeled $u$ (upper) and $l$ (lower). The circuit is shown in Fig.\ref{fig1}(b), where the qubits are labelled as $Q1$ and $Q2$ at the inputs and the orthogonal modes making up the qubits are shown as separate spatial rails. We note that it does not matter that the single photon and two photon components of the state end up in different spatio-temporal modes (at different average frequencies) because: (i) the coherent interactions that occur after the photon sorters only superpose modes containing two photon components, hence these interactions occur between matched modes; and (ii) in the end destructive measurements are made on all the modes that have been through the photon sorters.  If $\beta_{dir} < 1$ then there will be loss in the TLS and there will be heralded failure events when the photons do not make it through the circuit (see Supplementary Material for discussion and graph of probability of success). 
\begin{figure}[htb]
\begin{center}
\includegraphics*[width=8.5cm]{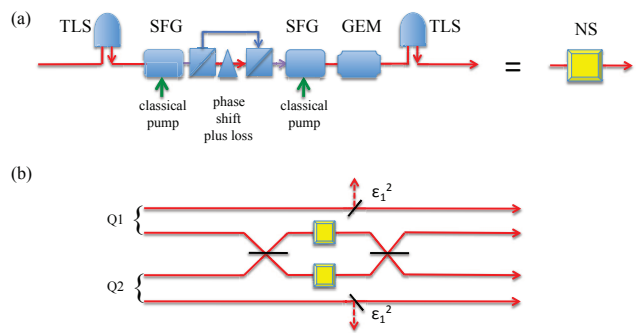}
\caption{Components of the non-linear sign (NS) gate and deterministic Controlled-Sign (CZ) gate: (a) the NS gate is constructed from a two-level scatterer (TLS) followed by Sum Frequency Generation (SFG), where the classical pump is in the mode $f'$. This is followed by a $\pi$ phase shift which is imposed only on the two photon term. If $\beta_{dir}<1$ then loss is also applied only to the two photon term. A second SFG then reverses the frequency shift, followed by short term storage in a gradient echo memory (GEM) which inverts the pulse shape. Finally the inverted pulse is sent through a second TLS which recombines the one and two photon terms back into the same mode. If the initial mode shape was time symmetric then the overall effect will be that of an NS gate, i.e. to impose a $\pi$ phase shift only on the two photon terms whilst leaving the mode shapes unchanged; (b) the resultant NS gates can be incorporated in a simple linear optical circuit to produce a CZ gate. Beamsplitters on the outer arms are required if $\beta_{dir} < 1$ in order to re-symmetrize the state.}
\label{fig2}
\end{center}
\end{figure}

{\it Deterministic CZ Gate}:
Given deterministic Bell measurements it is possible to construct a deterministic CZ gate using the techniques of gate teleportation \cite{GOT99} and linear optical quantum computing \cite{KNI01}. The necessary circuit is however quite complex requiring significant off-line optical resources for state preparation and hence either quantum memory or sophisticated real-time optical switching. In addition such a gate necessarily includes electro-optic feedforward.

It is interesting to ask if alternatively the TLS non-linearity plus Gaussian optics is sufficient to directly implement a deterministic CZ gate in an all-optical arrangement. In the following we show that this is possible in principle with TLS, SFG, gradient echo memory (GEM) and linear optics. The set-up is shown schematically in Fig.\ref{fig2}. We start with an input light field containing zero, one and two photon terms that can be written as:
\begin{equation}
\alpha |0 \rangle +\xi |1_f \rangle + \gamma |2_f \rangle
\label{start}
\end{equation}
As described in the previous section, the combination of TLS and SFG with a suitable classical pump leads to a state of the form:
\begin{equation}
\alpha |0 \rangle |0 \rangle_a +\xi |0 \rangle |1_{f'} \rangle_a + \gamma |\bar 2_{f'} \rangle |0 \rangle_a
\end{equation}
Because of the different frequencies of the one and two photon components they can be addressed individually, hence we impose a $\pi$ phase shift only on the two photon term -- in Fig.\ref{fig2} we represent this by spatially separating the beams, imposing the phase shift and then recombining them, but in practice easier techniques, such as using a wave-shaper, may be available. SFG is a reversible process, thus, by choosing a suitable phase relationship between the classical pump and the beams, the one photon component can be converted back to its original centre frequency. The output state after this manipulation is:
\begin{equation}
\alpha |0 \rangle +\xi |1_{f'} \rangle - \gamma |\bar 2_{f'} \rangle
\label{pregem}
\end{equation}
We now wish to undo the initial separation of the one and two photon terms into orthogonal modes by interacting with the TLS a second time. However, the mode distortion is not time-symmetric so we must first invert the pulse shape of the modes.  This can be achieved using a gradient echo memory (GEM) \cite{ALE06, HED10}. The GEM can be thought of as a material containing an ensemble of two-level atoms that can absorb and store an incident light pulse as it passes through it. During the storage or writing process a field gradient is applied to the material, producing a spatially selective storage of the
different frequency components of the input signal. To release or read out the pulse, the gradient is reversed and the light emerges from the other end of the material. However, the reversal of the gradient results in the shape of the pulse being inverted between input and output. In particular, in the limit that the storage bandwidth of the memory is much larger than the bandwidth of the pulse and the storage time of the memory is much longer than the pulse length, the action of the memory on an optical mode operator can be expressed as \cite{HUS13}:
\begin{eqnarray}
\hat a_{in}(t) &=& \int d k F(k) e^{-i k t} \hat a_{k}  \nonumber \\
%\hat a_{out}(t) &=&
& ^{GEM}_{\to}&  \int d k F(-k) e^{-i k (t-T)} \hat a_{k}
\label{gem}
\end{eqnarray}
The pulse is delayed by a time $T$ and the pulse shape is inverted. An explicit calculation confirms that if the state of Eq.\ref{pregem} is transformed according to Eq.\ref{gem} and then interacted a second time with a TLS, the final output is:
\begin{equation}
\alpha |0 \rangle +\xi |1_{f} \rangle - \gamma |2_{f} \rangle
\label{end}
\end{equation}
where we have assumed the initial mode shape was time symmetric. The total transformation from Eq.\ref{start} to Eq.\ref{end} is characteristic of a Nonlinear-Sign (NS) gate, as introduced in Ref.\cite{KNI01}. Two NS gates can be combined with linear optics to make a CZ gate as shown in Fig.\ref{fig2}(b). Consider first the case for which $\beta_{dir} =1$ and hence $\epsilon_1 = 1$. All but one of the possible two qubit logical input states lead to only zero or one photon occupation of the interferometer containing the NS gates in the central region of the circuit. The exception is the logical state with the lower rail of $Q1$ occupied and the upper rail of $Q2$ occupied. In this case, because of the Hong Ou Mandel effect \cite{HOM}, the {\it only} allowed photon arrangements in the central interferometer are a pair of photons through the upper NS gate or a pair of photons through the lower NS gate. Hence, only in this case a phase is imposed on the output state as required for a CZ gate. Notice that all the mode distortions are undone, hence a network of such gates may be used to implement universal quantum computation using single photon inputs. In Ref.\cite{KNI01} the NS gate was implemented with linear optics and had a probability of success of 25\%, hence leading to a CZ gate with probability of success 6.25\%. Here the NS gate and hence the CZ gate are in principle deterministic. In the case for which $\beta_{dir} < 1$ the gate will no longer be deterministic because photons can be lost in the TLS. In this case additional loss elements need to be introduced into the gate to ensure the qubit states do not become skewed (see Fig.\ref{fig2}(b) and Supplementary Material for discussion and plot of probability of success). 
%Fig.3 shows how the probability of success of the CZ gate changes as a function of $\beta_{dir}$ (where we assume unit efficiency for the SFG, GEM and detectors). 

{\it Discussion}:
We have shown that a deterministic Bell measurement and CZ gate can be implemented by combining a non-linear element with active and passive Gaussian optics. This is possible in spite of (or perhaps because of) the mode distortion produced by the non-linear element. We now discuss the challenges involved in implementing our schemes by briefly reviewing the state of the art for the various components.

Different platforms have been experimentally shown to be suitable for constructing a TLS \cite{LOD13}(see Fig.\ref{figcalcs}(a)). Single atoms and single quantum dots that are coupled to either photonic nanostructures \cite{LUK14, SEN14, KIMB14, ARC14} or whispering gallery mode resonators \cite{RAUSCH14, PAINT07}are the most promising at optical frequencies . Furthermore, transmon qubits in 1D transmission lines can be employed in the microwave regime \cite{HOI12}. In the case of quantum dots in photonic-crystal waveguides, coupling efficiencies of 98.4\% have been demonstrated in experiments on emission dynamics \cite{ARC14}. For the coherent scattering applications considered here any pure dephasing, spectral diffusion, and Raman scattering into the phonon-sideband will limit the performance. Excitingly $97 \%$ indistinguishability of single photons \cite{HE13} and about $95 \%$ of the emission in the zero-phonon line \cite{KONT12} have been experimentally reported. Mode selectivity of 80\%, with bandwidths compatible with quantum dot TLSs, has been experimentally demonstrated for SFG \cite{BRE14}, with excellent prospects for improvement. Hence, the technology required for implementing the Bell measurement protocol currently exists. The bottle neck in our CZ gate protocol is likely to be the GEM memories that, whilst showing good storage times and efficiency, currently operate with bandwidths around a MHz - and hence are currently incompatible with quantum dot bandwidths of around 320MHz. Never-the-less, there does not seem to be any in principle reasons why GEM of the necessary bandwidth could not be realized.

This research was supported by the Australian Research Council (ARC) under the Centre of
Excellence for Quantum Computation and Communication Technology (CE110001027). I.S., S.M. and P.L. gratefully acknowledge
financial support from the Danish Council for Independent
Research (Natural Sciences and Technology and Production
Sciences) and the European Research Council
(ERC consolidator grant ALLQUANTUM). 

\section{Supplementary Material}
\begin{figure}[htb]
\begin{center}
\includegraphics*[width=8.5cm]{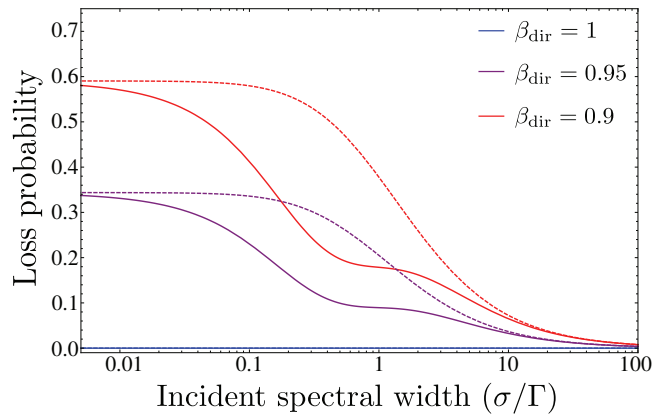}
\caption{Probability of loss in the TLS for two photon components (solid lines) and two single photons (dotted lines) as a function of $\beta_{dir}$. The non-linearity of the loss (for linear loss the dotted and solid curves would be the same) leads to the different probabilities of success for the Bell states and the skewing of the CZ-gate in Eq.\ref{loss4}. The fact that loss decreases with increasing spectral width means that the crossing points at higher spectral width in Fig.1(b) of the main text are the optimal choice.}
\label{figloss}
\end{center}
\end{figure}

In the main text we considered for the most part the case where $\gamma = 0$ ($\beta_{dir} = 1$). When $\gamma \ne 0$ ($\beta_{dir} < 1$) photons can be lost from the mode of interest resulting in a mixed state for the output mode.  However, we can continue to write the output state of the TLS as a pure state by formally including the lost modes in the output ket. Thus the output when a single photon state is incident on the TLS can be written
\begin{equation}
 |1_{f} \rangle \to  |1_{f'} \rangle + |not 1 \rangle
\label{loss1}
\end{equation}
where $|not 1 \rangle$ is an unnormalized composite ket containing all the state components for which a single photon has not been coupled into the output mode. It is useful to write the output in terms of normalized kets such that
\begin{equation}
 |1_{f} \rangle \to  \sqrt{\epsilon_1}|1_{f'} \rangle_N +  \sqrt{1-\epsilon_1}|not 1 \rangle_N
\label{loss1n}
\end{equation}
where $\epsilon_1 = \langle 1_{f'} |1_{f'} \rangle$ and in general $|\theta \rangle_N = |\theta \rangle/\sqrt{|\langle \theta|\theta \rangle|}$. For a two photon input we now find
\begin{eqnarray}
 |2_{f} \rangle &\to&  \epsilon_1(1-{{2 \eta}\over{\epsilon_1^2}})\; |2_{f'} \rangle_N + \sqrt{\epsilon_b - {{4 \eta^2}\over{\epsilon_1^2}}} \;\; |\bar 2_{f'} \rangle_N \nonumber \\
 &+& \sqrt{1-\epsilon_1^2+4 \eta -\epsilon_b}\;\;  |not 2 \rangle_N
\label{loss2}
\end{eqnarray}
where $\epsilon_b = \langle 2_{fb} |2_{fb} \rangle$ and now $|not 2 \rangle_N$ is a (normalized) ket containing all state components for which two photons are not coupled into the output mode. Analytic solutions can be found for the different probability amplitudes $$\epsilon_1=1-\frac{4 \gamma  \Gamma  (\gamma +\Gamma +4 \sigma )}{(\gamma +\Gamma ) (\gamma +\Gamma +2 \sigma )^2} $$ and $$\epsilon_b =\frac{16 \Gamma ^4 \sigma  \left(38 \sigma  (\gamma +\Gamma )+3 (\gamma +\Gamma )^2+96 \sigma^2\right)}{(\gamma +\Gamma )^2 (\gamma +\Gamma +2 \sigma )^3 (3 (\gamma +\Gamma )+2 \sigma) (\gamma +\Gamma +6 \sigma)},$$ where the incoming spectral amplitude of the photons is taken to be a Lorentzian of the form $$f_L(k)= \frac{\sqrt{\frac{2\sigma^3}{\pi}}}{\sigma ^2+(k-\text{k0})^2}.$$ The matching condition such that the two-photon component mode is orthogonal to the single photon component mode is now
\begin{equation}
\eta = {{\epsilon_1^2}\over{2}}
\label{loss2e}
\end{equation}
which tends to the lossless case of $0.5$ as $\epsilon_1$ tends to unity. Inserting this condition into Eq.\ref{loss2} gives
\begin{eqnarray}
 |2_{f} \rangle \to \sqrt{\epsilon_b - \epsilon_1^2} \;\; |\bar 2_{f'} \rangle_N %\nonumber \\
 + \sqrt{1+ \epsilon_1^2 -\epsilon_b}\;\;  |not 2 \rangle_N
\label{loss3}
\end{eqnarray}
In the limit of $\gamma \to 0$ we have $\epsilon_1 \to 1$ and (when the condition of Eq.\ref{loss2e} is fulfilled) $\epsilon_b \to 2$, and Eqs \ref{loss1n} and \ref{loss3} reduce to the lossless cases in the main text. The behaviour of the losses as a function of the spectral width is shown in Fig.\ref{figloss}.

\begin{figure}[htb]
\begin{center}
\includegraphics*[width=8.5cm]{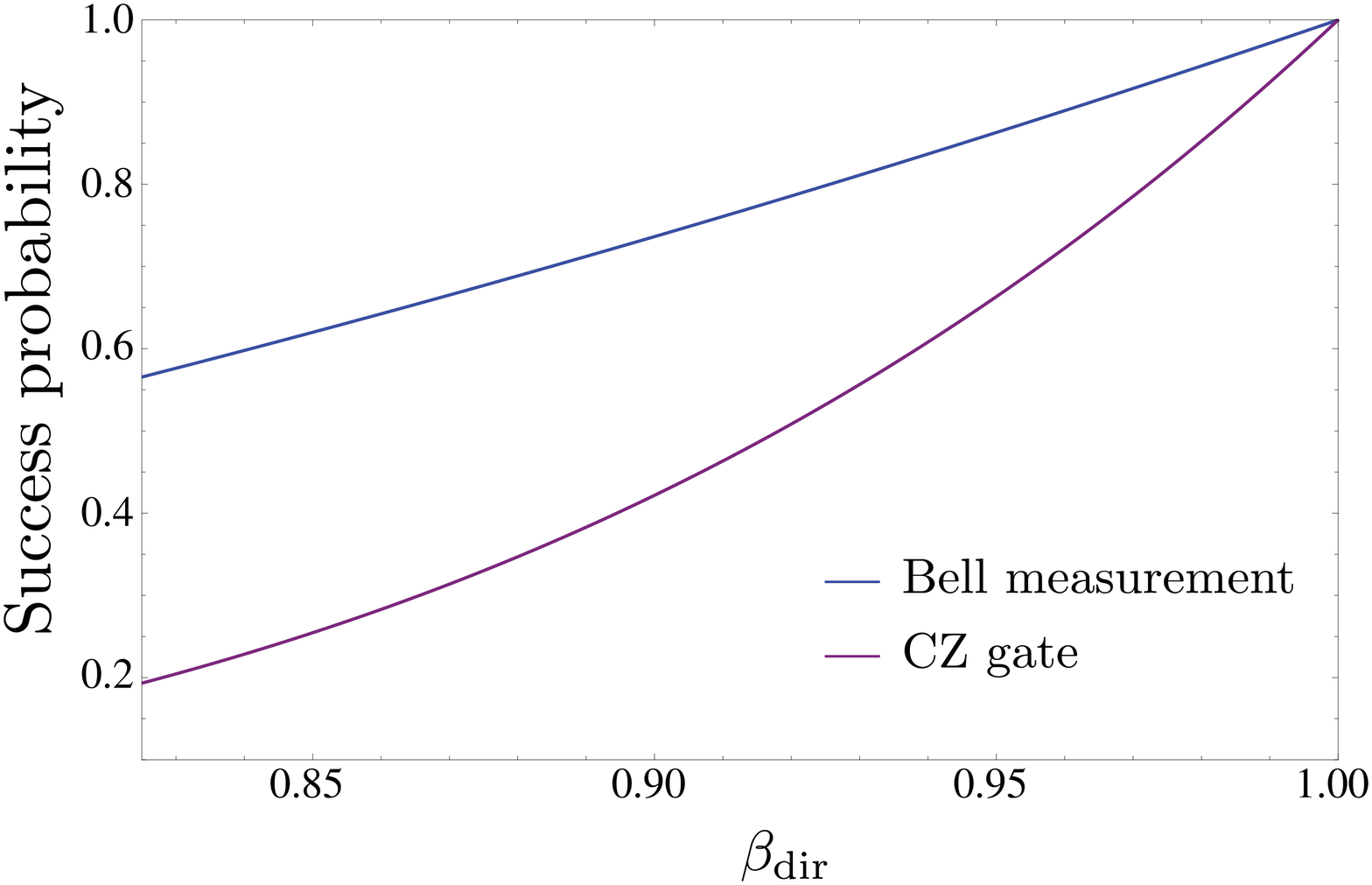}
\caption{Probability of success of the Bell Measurement (upper curve) and the CZ gate (lower curve) as a function of $\beta_{dir}$, where the SFG, GEM, detectors and other linear optical elements have been assumed to have unit efficiency.}
\label{fig3}
\end{center}
\end{figure}

If the TLS are used to make a Bell measurement, the effect of the loss is to lead to a non-unity probability of success. In order for $|\psi^{\pm} \rangle$ to be identified two single photon scattering events must succeed, thus the probability of success will be $\epsilon_1^2$. On the other hand to identify $|\phi^{\pm} \rangle$ requires a single two-photon scattering event to succeed, thus the probability of success will be $(\epsilon_b - \epsilon_1^2)$, leading to an average probability of success of $\epsilon_b/2$. Unsuccessful events are heralded by the failure to detect two photons. In Fig.\ref{fig3} we plot the probability of success of the Bell measurement versus the $\beta_{dir}$ factor (where we assume unit efficiency for the SFG and detectors).

We now consider the CZ gate. The loss introduced by the four TLS affect the various state components differently producing  a skewing of the output state. This can be compensated to some extent by inserting additional linear loss of transmission $\epsilon_1$ on the outer arms of the CZ gate (see Fig.3(b) main text). The state transformation of the entire gate then becomes
\begin{eqnarray}
&& \alpha \; |0 , 1_{f} \rangle |0 , 1_{f} \rangle + \xi \; |1_{f} , 0 \rangle |1_{f} , 0 \rangle + \gamma \;|1_{f} , 0 \rangle |0 , 1_{f} \rangle + \nonumber \\
&& \;\;\;\;\; \delta \;|0 , 1_{f} \rangle |1_{f} , 0 \rangle \;\;\;\;\; \to \nonumber \\
&& \epsilon_1^2 (\alpha \;|0 , 1_{f} \rangle |0 , 1_{f} \rangle + \xi  \;|1_{f} , 0 \rangle |1_{f} , 0 \rangle + \gamma \;|1_{f} , 0 \rangle |0 , 1_{f} \rangle) - \nonumber \\
&& \;\;\;\;\; (\epsilon_b - \epsilon_1^2)\; \delta \;|0 , 1_{f} \rangle |1_{f} , 0 \rangle +  |not 2 \rangle
\label{loss4}
\end{eqnarray}
where we have used the short hand $|0 , 1_{f} \rangle = |0 \rangle |1_{f} \rangle$ to represent the logical zero state component of a dual rail qubit with no photon in the top mode and one in the lower mode, and similarly for the logical one state component, $|1_{f}, 0 \rangle = |1_{f} \rangle |0 \rangle$. The first ket in Eq.\ref{loss4} represents the first qubit ($Q1$) and the second ket, the second qubit ($Q2$). The state component $|not 2 \rangle$ is a an unnormalized composite ket representing all state components for which the total photon number coupled into the four output modes is less than two photons. Again, in the limit of $\gamma \to 0$ we have $\epsilon_1 \to 1$ and $\epsilon_b \to 2$, and the gate becomes an ideal CZ gate. Two different kinds of errors can occur when $\gamma \ne 0$. The first are locatable errors that arise due to the non-zero amplitude of the $ |not 2 \rangle$ component. These are locatable because they can be identified by an incorrect number of photons being detected at the end of a calculation or alternatively via non-destructive number measurement during a calculation. In contrast the fact that in general $(\epsilon_b - \epsilon_1^2) \ne \epsilon_1^2$ means that there is a skewing of the output state that can lead to non-locatable or logical errors in any computation, that would require a full error correction code to correct. Fortunately the skewing can be easily corrected by introducing some additional loss, $\eta_2$, along with the phase shift on the 2 photon component in the middle of the NS gate (Fig.3(b) main text). If the value of this loss is chosen such that $\eta_2 = {{\epsilon_1^2}\over{\epsilon_b - \epsilon_1^2}}$ then Eq.\ref{loss4} becomes

\begin{eqnarray}
&& \alpha \; |0 , 1_{f} \rangle |0 , 1_{f} \rangle + \xi \; |1_{f} , 0 \rangle |1_{f} , 0 \rangle + \gamma \;|1_{f} , 0 \rangle |0 , 1_{f} \rangle + \nonumber \\
&& \;\;\;\;\; \delta \;|0 , 1_{f} \rangle |1_{f} , 0 \rangle \;\;\;\;\; \to \nonumber \\
&& \epsilon_1^2 (\alpha \;|0 , 1_{f} \rangle |0 , 1_{f} \rangle + \xi  \;|1_{f} , 0 \rangle |1_{f} , 0 \rangle + \gamma \;|1_{f} , 0 \rangle |0 , 1_{f} \rangle - \nonumber \\
&& \;\;\;\;\; \; \delta \;|0 , 1_{f} \rangle |1_{f} , 0 \rangle) +  |not 2 \rangle
\label{loss5}
\end{eqnarray}
hence the skewing is removed and the overall probability of success of the CZ gate becomes $\epsilon_1^4$.

\end{document}